\documentclass[
aps,prd,
12pt,
nopreprintnumbers,
showpacs,
eqsecnum,
nofootinbib
]{revtex4-1}

\usepackage{graphicx}
\usepackage{amssymb}

\begin{document}

\title{Magnetic monopoles in multi-vector boson theories
}
\author{Koichiro Kobayashi}\email[]{kobayasi@oshima-k.ac.jp}
\affiliation{National Institute of Technology, Oshima College,\\
Suooshima-cho, Yamaguchi 742--2193, Japan
}
\author{Nahomi Kan}\email[]{kan@gifu-nct.ac.jp}
\affiliation{National Institute of Technology, Gifu College,\\
Motosu-shi, Gifu 501--0495, Japan
}
\author{Kiyoshi Shiraishi}\email[]{shiraish@yamaguchi-u.ac.jp}
\affiliation{
Graduate School of Sciences and Technology for Innovation, Yamaguchi
University, 
Yamaguchi-shi, Yamaguchi 753--8512, Japan}
\date{\today}

\begin{abstract}
A classical solution for a magnetic monopole is found in a specific multi-vector
boson theory.
We consider the model whose $[SU(2)]^{N+1}$ gauge group is broken by
sigma-model fields (\`a la dimensional deconstruction) and further
spontaneously broken by an adjoint scalar (\`a la triplet Higgs mechanism).
In this multi-vector boson theory, we find the solution for the monopole whose mass
is $M_N\sim\frac{4\pi v}{g}\sqrt{N+1}$, where $g$ is the common gauge coupling
constant and $v$ is the vacuum expectation value of the triplet Higgs field,
by using a variational method with the simplest set of test functions. 
\end{abstract}



\maketitle

\section{Introduction}
\label{sec1}

The existence of magnetic monopoles
(for reviews, see \cite{GO,Rajaraman,Rubakov,MS,Shnir,Weinberg}) has been
discussed for many years, although monopoles have not yet been observed
experimentally.

In 1931, Dirac \cite{Dirac} reconsidered the duality in electromagnetism and showed
that the quantum mechanics of an electrically charged particle can be consistently
formulated in the presence of a point magnetic charge, provided that the magnetic
charge $g_m$ is related to the electric charge $e$ by $eg_m=n\hbar c/2$ with an
integer
$n$.
In 1974, `t Hooft \cite{tH} and Polyakov \cite{Po} found that a nonsingular
configuration arises from spontaneous symmetry breaking in a certain class of
non-Abelian gauge theory.
Their models are based on the Georgi-Glashow model
\cite{GG}, which uses spontaneous symmetry breaking of $SU(2)$ gauge symmetry by a
scalar field in the adjoint representation.
The `t Hooft--Polyakov monopoles are classical solutions, which are stable for
topological reasons. Recently, the mathematical study of monopoles have focused on
not only topology, but also integrable systems, supersymmetry, nonperturbative
analyses, and so on.

In the present paper, we consider a novel monopole in a multi-vector boson theory,
which is based on dimensional deconstruction \cite{ACG,HPW} and the Higgsless
theories
\cite{CHPW,CGMPT,CGPT,CHM,FGS,Georgi}.
The Higgsless theory is one of the theories that
include symmetry breaking of the electroweak symmetry. In the Higgsless
theory, for example, the
$[SU(2)]^N\otimes U(1)$ gauge theory is considered. Such a theory yields
$N$ sets of massive vector fields besides one massless photon field.

In our model of the multi-vector boson theory,  $[SU(2)]^{N+1}$ gauge symmetry is
assumed.  One of the $SU(2) $ gauge groups is broken by an adjoint scalar as
in the Georgi-Glashow model. 
There remains one massless vector field due to the triplet Higgs mechanism. 
We can thus construct the `t Hooft--Polyakov-type monopole configuration in the
model.
We estimate the monopole mass $M\sim\frac{4\pi v}{g}\sqrt{N+1}$, where $v$ is the
vacuum expectation value of the scalar field, and $g$ is the coupling constant of
the gauge field. 

In Sec.~\ref{sec2}, we briefly review dimensional deconstruction and the
Higgsless theory.
Our model of the multi-vector boson theory is shown in Sec.~\ref{sec3},
which is a generalization of the gauge-field part of the Higgsless theory.
The mass spectrum in the multi-vector boson theories is investigated in
Sec.~\ref{sec4}. In Sec.~\ref{sec5}, we demonstrate the construction of monopole
configurations in the multi-vector boson theory. In order to treat many
variables, we propose an approximation scheme by a variational method in this
section. In Sec.~\ref{sec6}, we discuss the magnetic charge of the monopole in the
multi-vector boson theory. The final section (Sec.~\ref{sd}) is devoted to summary
and discussion. 


\section{Deconstruction and Higgsless theory}
\label{sec2}

We review the basic idea of dimensional deconstruction \cite{ACG,HPW} and the
Higgsless theories \cite{CHPW,CGMPT,CGPT,CHM,FGS,Georgi}
in this section.
We consider $N+1$ gauge fields $A_{1\mu}$, $A_{2\mu}$, \dots $A_{N+1,\mu}$.
The field strength $G_{I\mu\nu}$ ($I=1,2,\dots, N+1$) is defined as
\begin{equation}
G_{I\mu\nu}\equiv\partial_\mu A_{I\nu}-\partial_\nu A_{I\mu}-ig_I[
A_{I\mu}, A_{I\nu}]\,,
\end{equation}
where $g_I$ is the $I$-th gauge coupling constant.
The $I$-th field strength transforms as
\begin{equation}
G_{I\mu\nu}\rightarrow U_IG_{I\mu\nu} U_{I}^\dagger\quad
(1\le I\le N+1)\,,
\end{equation}
according to the $I$-th gauge group transformation $U_I\in G_I$.

In addition to the gauge fields, we introduce $N$ scalar fields 
$\Sigma_{1}$, $\Sigma_{2}$, \dots $\Sigma_{N}$, which would supply
the Nambu-Goldstone fields as non-linear-sigma-model fields.
The scalar field $\Sigma_I$ ($I=1,2,\dots, N$) transforms as in the
bi-fundamental representation,
\begin{equation}
\Sigma_I\rightarrow U_I\Sigma_I U_{I+1}^\dagger\quad(1\le I\le N)\,.
\end{equation}
(Here, we show the case of `linear moose', and the different assignments of the
transformation of $\Sigma_I$ yield the theory associated with various other
types of moose diagrams
\cite{ACG,HPW,CHPW,CGMPT,CGPT,CHM,FGS,Georgi}.)

Now, the Lagrangian density, which is invariant under the gauge transformation of
$G_1\otimes G_2\otimes\cdots\otimes G_{N+1}$, is given by
\begin{equation}
\mathcal{L}=-\frac{1}{2}\sum_{I=1}^{N+1}\mathrm{Tr}\, G_{I\mu\nu}G_I^{\mu\nu}
-\sum_{I=1}^{N}\mathrm{Tr}\, (D_\mu\Sigma_{I})^\dagger(D^\mu\Sigma_{I})\,,
\end{equation}
where the covariant derivative of $\Sigma_I$ is
\begin{equation}
D_\mu\Sigma_I\equiv\partial_\mu\Sigma_I-ig_IA_{I\mu}\Sigma_I
+ig_{I+1}\Sigma_IA_{I+1,\mu}\,,
\end{equation}
and then its gauge transformation is 
\begin{equation}
D_\mu\Sigma_I\rightarrow U_I(D_{\mu}\Sigma_I) U_{I+1}^\dagger\,.
\end{equation}

In the usual dimensional deconstruction scheme,
we consider that $G_1=G_2=\cdots=G_{N+1}=G$ and $g_1=g_2=\cdots=g_{N+1}=g$.
We also assume that the absolute value of each non-linear sigma model field
$|\Sigma_I|$ has a common vacuum value, $f$.
Then, the field $\Sigma_I$ is expressed as 
\begin{equation}
\Sigma_I=f\exp\left(i\frac{\pi^aT^a}{f}\right)\,,
\end{equation}
where $T^a$ is the generator in the adjoint representation of $G$, and $\pi^a$ is
the Nambu-Goldstone field, which is absorbed into the gauge fields.
Taking the unitary gauge $\Sigma_I=f\times \mbox{(identity matrix)}$, we find that
the kinetic terms of $\Sigma_I$ lead to the mass terms of the gauge fields as
(provided that  $\mathrm{Tr}\,(T^aT^b)=\frac{1}{2}\delta^{ab}$)
\begin{equation}
\sum_{I=1}^{N}\mathrm{Tr}\, (D_\mu\Sigma_{I})^\dagger(D^\mu\Sigma_{I})=
\frac{1}{2}g^2f^2\sum_{I=1}^{N}(A_{I\mu}^a-A_{I+1,\mu}^a)^2\,,
\label{mt}
\end{equation}
and these produce the mass spectrum of vector bosons. 
It is known that a certain continuum limit of this model can be taken, which
corresponds to the $G$ gauge theory with one-dimensional compactification
on to $S^1/Z_2$ (or an `interval').

In the Higgsless theories, for example, the gauge group $[SU(2)]^N\otimes U(1)$ is
adopted for explaining the electroweak sector in the particle theory.
Namely, we set $G_1=U(1)$ and $G_2=\cdots=G_{N+1}=SU(2)$.
Then, the covariant derivative of $\Sigma_1$ is
\begin{equation}
D_\mu\Sigma_1\equiv\partial_\mu\Sigma_1-ig_1A_{1\mu}T^3\Sigma_1
+ig\Sigma_1A_{2\mu}\,,
\end{equation}
where $g_1$ is the $U(1)$ gauge coupling constant, $g$ is the common
$SU(2)$ gauge coupling constant, $A_{1\mu}$ is the
$U(1)$ gauge field, and $T^3$ is the third generator of $SU(2)$.
The non-zero vacuum expectation value of $\Sigma_I$ leads to
symmetry breaking $[SU(2)]^N\otimes U(1)\rightarrow U(1)$
\cite{CHPW,CGMPT,CGPT,CHM,FGS,Georgi}, and we get only one massless electromagnetic
field and $N$ sets of massive weak boson fields.

The original motivation for the Higgsless theory has been
abandoned after the discovery of the Higgs particles. 
Nevertheless, we would like to extend the standard model, since there might be a
lack of unknown extra particles, which explain the dark matter problem
\cite{BHS,Profume}.
As a model of dark matter, the multi-vector boson theory describes a hidden sector
of dark photons \cite{Okun,Holdom} with mutual mixings. 
Therefore, we suppose that it is
worth considering the theoretical models whose massive particle contents are rich
and governed by certain symmetries.


\section{multi-vector boson theory from the Higgsless theory incorporating the
Higgs mechanism}
\label{sec3}

Here, we consider the model whose $[SU(2)]^N\otimes U(1)$ gauge group comes
from the spontaneous symmetry breaking by an adjoint scalar \cite{GG}: 
$[SU(2)]^{N+1}\rightarrow[SU(2)]^{N}\otimes U(1)$. 
The mechanism is now generally called the Higgs mechanism.
The symmetry is broken into $U(1)$ by the vacuum expectation value of the
non-linear sigma model field $\Sigma_I$ introduced in the previous section. 
As a consequence, we have a monopole configuration; the construction of the
monopole solution will be described in the next section.
In this section, we define our model, and in the subsequent section, we show the
mass spectrum of this model.

We consider the following Lagrangian density:
\begin{equation}
\mathcal{L}=-\frac{1}{2}\sum_{I=1}^{N+1}\mathrm{Tr}\, G_{I\mu\nu}G_I^{\mu\nu}
-\sum_{I=1}^{N}\mathrm{Tr}\, (D_\mu\Sigma_{I})^\dagger(D^\mu\Sigma_{I})
-\mathrm{Tr}\, (D_\mu\phi)^\dagger(D^\mu\phi)
-\frac{\lambda}{4}(2\mathrm{Tr}\,\phi^\dagger\phi-v^2)^2\,,
\label{3LD}
\end{equation}
where $G_{I\mu\nu}$ $(I=1,\dots,N+1)$ is the field strength of the $SU(2)_I$ gauge
field
$A_{I\mu}$ $(I=1,\dots,N+1)$, and
$\Sigma_I$ $(I=1,\dots,N)$ is the non-linear sigma model fields in the
bi-fundamental representation of $SU(2)_I\otimes SU(2)_{I+1}$, which connect the
gauge fields at neighboring sites, as in the dimensionally deconstructed model
reviewed in the previous section. For simplicity, all the coupling constants of the
gauge fields are assumed to be the same $g$.

Here, $\phi$ is a scalar field in the adjoint representation of $SU(2)_1$, and the
covariant derivative of the scalar field $\phi$ is given by
\begin{equation}
D_\mu\phi\equiv\partial_\mu\phi-ig[A_{1\mu}, \phi]\,.
\end{equation}
In the last term in the Lagrangian density (\ref{3LD}), $\lambda$ is a positive
constant and the constant
$v$ is the scalar field vacuum expectation value.

First, we consider the symmetry breaking by the sigma fields.
We choose the unitary gauge $\Sigma_1=\cdots=\Sigma_{N}=f\times
\mbox{(the identity matrix)}$. Then,
the Lagrangian density is represented as follows:
\begin{equation}
\mathcal{L}=-\frac{1}{4}\sum_{I=1}^{N+1}G_{I\mu\nu}^aG_I^{a\mu\nu}
-\frac{1}{2}g^2f^2\sum_{I=1}^{N}(A_{I\mu}^a-A_{I+1,\mu}^a)^2
-\frac{1}{2}\, (D_\mu\phi^a)(D^\mu\phi^a)
-\frac{\lambda}{4}(\phi^a\phi^a-v^2)^2\,,
\end{equation}
where
\begin{equation}
G_{I\mu\nu}^a\equiv \partial_\mu A_{I\nu}^a-\partial_\nu A_{I\mu}^a
+g\varepsilon^{abc}A_{I\mu}^bA_{I\nu}^c\quad\mbox{and}\quad
D_{\mu}\phi^a\equiv \partial_\mu \phi^a
+g\varepsilon^{abc}A_{1\mu}^b\phi^c\,.
\end{equation}
Here, we use the component representations $A_{I\mu}=A_{I\mu}^aT^a$,
$G_{I\mu\nu}=G_{I\mu\nu}^aT^a$, $\phi=\phi^aT^a$, and
$D_{\mu}\phi=D_{\mu}\phi^aT^a$, and $\varepsilon^{abc}$ is the totally
antisymmetric symbol ($a=1, 2, 3$).

Next, we consider the symmetry breakdown by the Higgs mechanism with respect to
the adjoint scalar field $\phi$.
We express the third component of the scalar field as $\phi^3=v+\varphi$.
Then, the Lagrangian density is denoted by
\begin{eqnarray}
\mathcal{L}&=&-\frac{1}{4}\sum_{I=1}^{N+1}G_{I\mu\nu}^1G_I^{1\mu\nu}
-\frac{1}{2}g^2f^2\sum_{I=1}^{N}(A_{I\mu}^1-A_{I+1,\mu}^1)^2
-\frac{1}{2}g^2v^2A_{1\mu}^1A_1^{1\mu}
\nonumber \\
& &-\frac{1}{4}\sum_{I=1}^{N+1}G_{I\mu\nu}^2G_I^{2\mu\nu}
-\frac{1}{2}g^2f^2\sum_{I=1}^{N}(A_{I\mu}^2-A_{I+1,\mu}^2)^2
-\frac{1}{2}g^2v^2A_{1\mu}^2A_1^{2\mu}
\nonumber \\
& &-\frac{1}{4}\sum_{I=1}^{N+1}G_{I\mu\nu}^3G_I^{3\mu\nu}
-\frac{1}{2}g^2f^2\sum_{I=1}^N(A_{I\mu}^3-A_{I+1,\mu}^3)^2
\nonumber \\
& &
-\frac{1}{2}\partial_\mu\varphi\partial^\mu\varphi-\lambda v^2\varphi^2
+\mbox{(interaction terms)}\,,
\label{LD}
\end{eqnarray}
where the labels $a$ are explicitly represented.
We have only one massless $U(1)$ symmetric gauge field in the third component.
Therefore, we have obtained the symmetry breaking $SU(2)\rightarrow
U(1)$ by using the Higgs mechanism. 
This type of symmetry breaking gives rise to the `t Hooft--Polyakov monopole
configuration. 

It should be noted that we do not discuss which sequences of symmetry breaking,
that is, 
$[SU(2)]^{N+1}\rightarrow [SU(2)]^{N}\otimes U(1)\rightarrow U(1)$
or $[SU(2)]^{N+1}\rightarrow SU(2)\rightarrow U(1)$ occurred in the universe,
although the order may have an effect on the process of creation of monopoles
in the early universe.


\section{Mass spectrum of vector bosons}
\label{sec4}

In the Lagrangian density (\ref{LD}), the mass term of gauge fields for $a=1$ is
\begin{eqnarray}
& &\mathcal{L}_{mass~term~a=1}=
-\frac{1}{2}g^2f^2\sum_{I=1}^{N}(A_{I\mu}^1-A_{I+1,\mu}^1)^2
-\frac{1}{2}g^2v^2A_{1\mu}^1A_1^{1\mu}\nonumber \\
& &=-\frac{1}{2}g^2f^2 \nonumber \\
& &\times
\left(A_1^{1\mu}~A_2^{1\mu}~A_3^{1\mu}~\cdots~A_{N-1}^{1\mu}~A_N^{1\mu}~A_{N+1}^{\mu}\right)
\left(
\begin{array}{rrrrrrr}
1+\frac{v^2}{f^2} & -1 & &&&& \\
-1 & 2 & -1 &&&& \\
  & -1 & 2 &-1&&&\\
&&\ddots&\ddots &\ddots&& \\
&&&-1& 2 &-1& \\
&&&&-1 & 2 & -1 \\
&&&& & -1& 1
\end{array}
\right)
\left(
\begin{array}{c}
A_{1\mu}^1 \\
A_{2\mu}^1 \\
A_{3\mu}^1\\
\vdots \\
A_{N-1,\mu}^1 \\
A_{N\mu}^1 \\
A_{N+1,\mu}^1
\end{array}
\right)\,.\nonumber\\
\end{eqnarray}
Therefore, for $a=1$, the mass-squared matrix $(mass^1)^2$ of the vector bosons
is
\begin{equation}
(mass^1)^2=g^2f^2\left(
\begin{array}{ccccccc}
1+\frac{v^2}{f^2} & -1 & &&&& \\
-1 & 2 & -1 &&&& \\
  & -1 & 2 &-1&&&\\
&&\ddots&\ddots &\ddots&& \\
&&&-1& 2 &-1& \\
&&&&-1 & 2 & -1 \\
&&&& & -1& 1
\end{array}
\right)\equiv g^2f^2(M^1)^2\,.
\end{equation}

We consider the eigenvalue equation
\begin{equation}
(M^1)^2\mathbf{A}^1=(M^1_E)^2\mathbf{A}^1\,,
\end{equation}
where $\mathbf{A}^1$ is the eigenvector
\begin{equation}
\mathbf{A}^1\equiv\left(
\begin{array}{c}
A_{1\mu}^1 \\
A_{2\mu}^1 \\
A_{3\mu}^1\\
\vdots \\
A_{N-1,\mu}^1 \\
A_{N\mu}^1 \\
A_{N+1,\mu}^1
\end{array}
\right)\,,
\end{equation}
and $(M_E^1)^2$ is the eigenvalue.

\begin{figure}[ht]
\centering
\includegraphics[width=7cm]{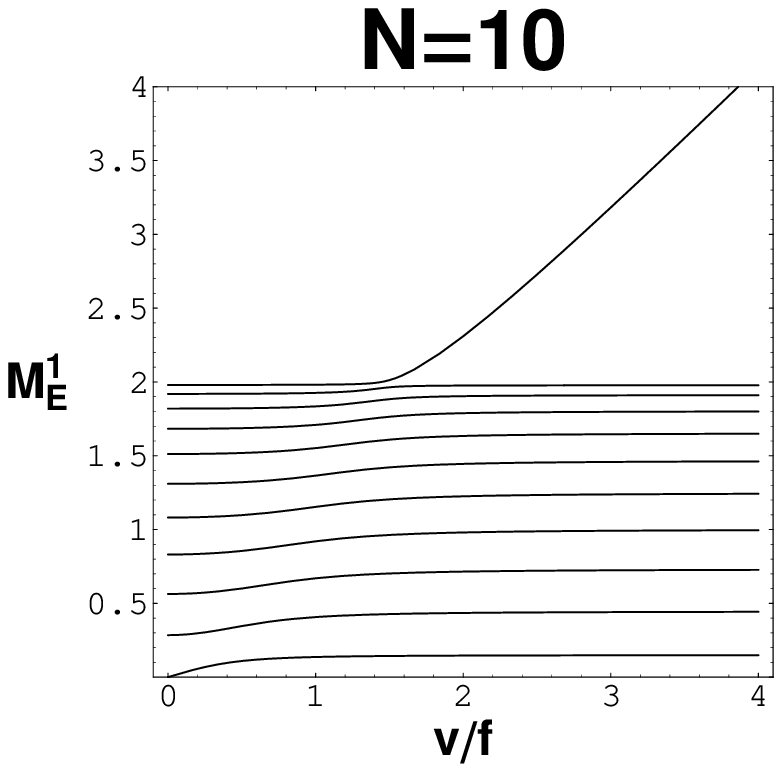}
\includegraphics[width=7cm]{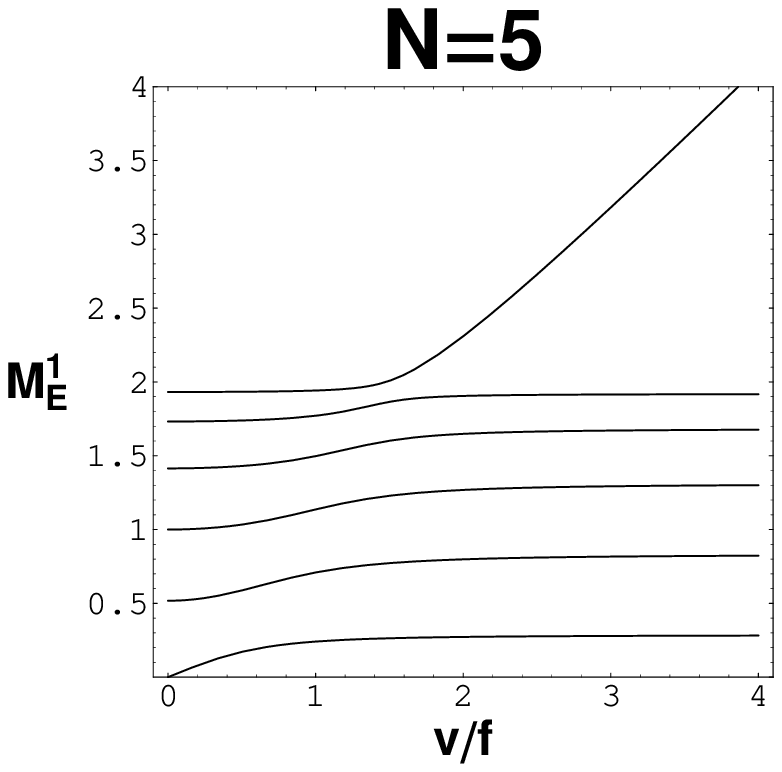}\\
\includegraphics[width=7cm]{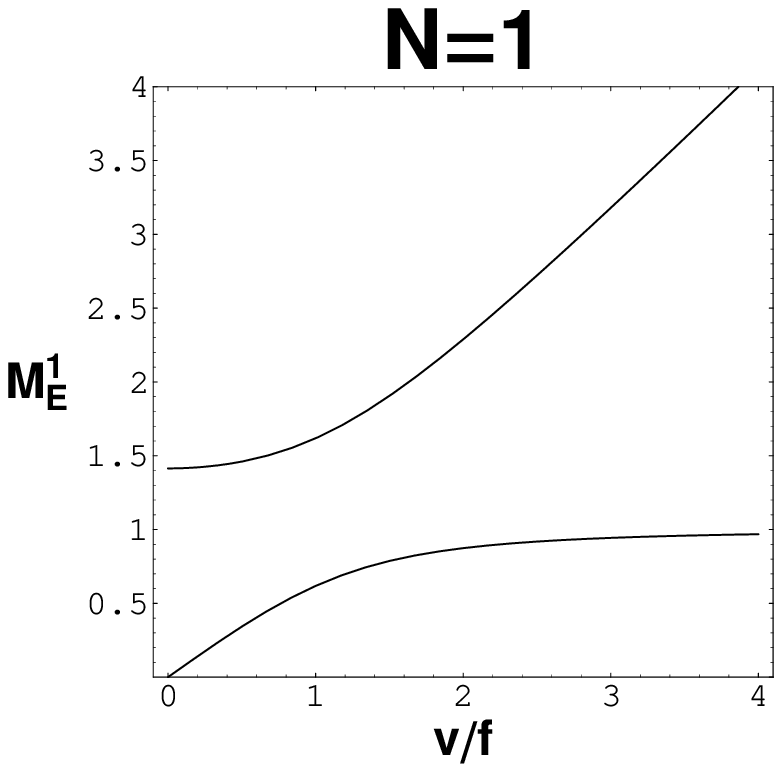}
\includegraphics[width=7cm]{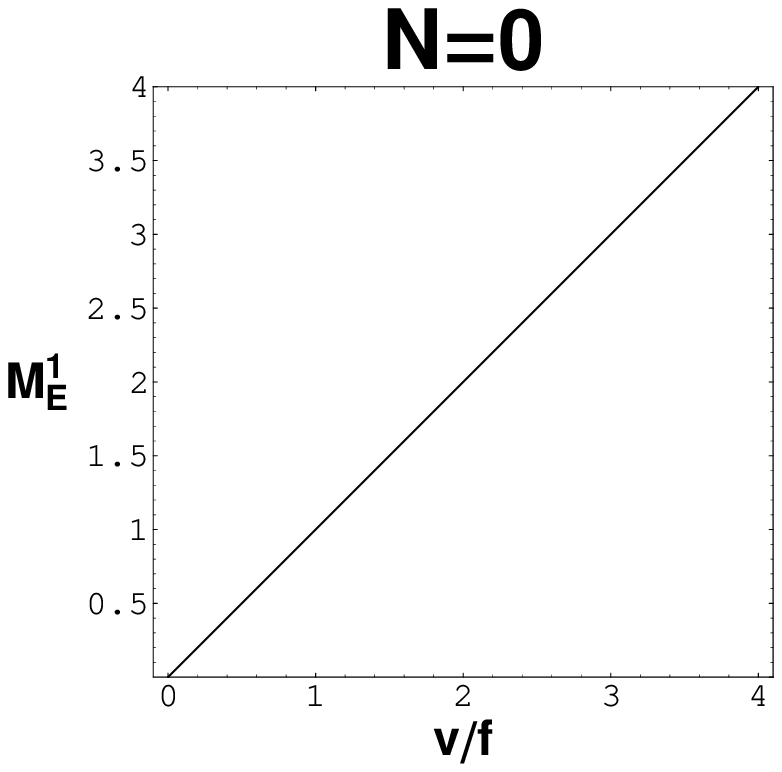}
\caption{The eigenvalues $M_E^1$ are shown as functions of $v/f$ in the cases of
$N=10, 5, 1,
\mbox{and~} 0$. }
\label{fig1}
\end{figure}

We show the $M_E^1$-$v/f$ graphs in Fig.~\ref{fig1}. The highest
eigenvalue behaves differently from the other eigenvalues.
When $v/f\rightarrow\infty$, the highest eigenvalue becomes $M_E^1\sim v/f$, but
the other eigenvalues asymptotically approach constant values that are less
than two.

The mass term of gauge fields for $a=2$ is the same as for $a=1$, but 
the mass term is different for $a=3$. The mass-squared matrix of gauge fields for
$a=3$ is
\begin{equation}
(mass^3)^2=g^2f^2\left(
\begin{array}{ccccccc}
1 & -1 & &&&& \\
-1 & 2 & -1 &&&& \\
  & -1 & 2 &-1&&&\\
&&\ddots&\ddots &\ddots&& \\
&&&-1& 2 &-1& \\
&&&&-1 & 2 & -1 \\
&&&& & -1& 1
\end{array}
\right)\equiv g^2f^2(M^3)^2\,.
\end{equation}

The eigenvalues $M_E^3$ can be analytically obtained \cite{CHPW} as
\begin{equation}
(M_E^3)_n=2\sin\frac{n\pi}{2(N+1)}\quad (n=0,\dots,N)\,.
\end{equation}
Obviously, there is a zero mode, and we have only one massless vector field in the
theory after symmetry breakdown.

\section{Energy and equations of motion of the monopole in the
multi-vector boson theory}
\label{sec5}

In the multi-vector boson theory defined by the Lagrangian density (\ref{LD}),
the `t Hooft--Polyakov-type monopole is expected.


Similar to the `t Hooft--Polyakov monopole, the static and spherically
symmetric monopole solution in the multi-vector boson theory is considered to be
specified by the following ansatz
\begin{eqnarray}
\phi^a&=&\delta_{ia}\frac{x^i}{gr^2}H(r)\,,\\
A_{I0}^a&=&0\,,\\
A_{Ii}^a&=&\varepsilon_{aij}\frac{x^j}{gr^2}[1-K_I(r)]\,,
\end{eqnarray}
and the boundary conditions on the function of the radial coordinate $r$
are
\begin{equation}
\lim_{r\rightarrow 0}H(r)/r=0\,,\quad
\lim_{r\rightarrow \infty}H(r)/r=gv\,,\quad
\lim_{r\rightarrow 0}K_I(r)=1\,,\quad
\lim_{r\rightarrow \infty}K_I(r)=0\,.
\end{equation}
The common form of $A_{Ii}^a$ is due to the requirement of finite energy of the
monopole, i.e., the contribution of the term (\ref{mt}) to the energy density
vanishes at spatial infinity.

For the static case, the energy density is given by $-\mathcal{L}$.
Substituting the ansatz, we obtain the expression for total energy
\begin{eqnarray}
& &E_N=\frac{4\pi v}{g}\int_0^\infty d\xi\left[\sum_{I=1}^{N+1}
\left\{(K_I')^2+\frac{1}{2\xi^2}(1-K_I^2)^2\right\}\right.\nonumber \\
& &\qquad+\left.\frac{f^2}{v^2}\sum_{I=1}^N(K_I-K_{I+1})^2+\frac{1}{2}
\left(H'-\frac{H}{\xi}\right)^2+\frac{1}{\xi^2}K_1^2H^2+
\frac{\lambda\xi^2}{4g^2}\left(\frac{H^2}{\xi^2}-1\right)^2\right]\,,
\end{eqnarray}
where we set $\xi\equiv gvr$ and the prime (${}'$) denotes the derivative with
respect to $\xi$.

From this expression, we can obtain the following equations of motion
by the variational principle:
\begin{eqnarray}
\xi^2K_1''&=&K_1(K_1^2-1)+H^2K_1+\frac{f^2}{v^2}\xi^2(K_1-K_2)\,,\\
\xi^2K_I''&=&K_I(K_I^2-1)+\frac{f^2}{v^2}\xi^2(2K_I-K_{I-1}-K_{I+1})\quad
(2\le I\le N)\,,\\
\xi^2K_{N+1}''&=&K_{N+1}(K_{N+1}^2-1)+\frac{f^2}{v^2}\xi^2(K_{N+1}-K_{N})\,,\\
\xi^2H''&=&2HK_1+\frac{\lambda}{g^2} H\left({H^2}-\xi^2\right)\,.
\end{eqnarray}

Analytical and semi-analytical studies of the single `t Hooft--Polyakov monopole
are found in Refs.~\cite{KZ,Gardner,FOR}.
Because it is hard 
 to find a set of solutions for these coupled equations for
large
$N$, and because we are presently considering a simple toy model, we adopt a
simple variational method to obtain approximate solutions in this paper.
We have confirmed that this approach obtains a good solution for the `t
Hooft--Polyakov monopole in the BPS limit.


For the approximation, we
assume that the solutions take the following forms:
\begin{eqnarray}
K_I(\xi)&=&(1+a_I\xi)\exp(-a_I\xi)\quad (1\le I\le N+1)\,,
\label{k1}\\
\frac{H(\xi)}{\xi}&=&1-\exp(-\alpha\xi)
\label{h1}\,,
\end{eqnarray}
where both $a_I$ $(1\le I\le N+1)$ and $\alpha$ are variational parameters.
The functions $K_I(\xi)$ and $H(\xi)$ with minimal number of parameters
apparently satisfy the boundary conditions and are similar to those of the
solutions in the `t Hooft--Polyakov monopole.
This is the reason why we assume the simple form of solutions as shown above.

We substitute the expressions (\ref{k1}) and (\ref{h1})
into the energy $E_N$, and calculate the minimum value of the energy
$E_N$ by varying the parameters $a_I$ and $\alpha$.

Each term is separately integrated as follows.
\begin{eqnarray}
& &\int_0^\infty d\xi
\left\{(K_I')^2+\frac{1}{2\xi^2}(1-K_I^2)^2\right\}=
\frac{41}{64}a_I\,, \\
& &\int_0^\infty d\xi
(K_I-K_{I+1})^2=
\frac{5}{4a_I}+\frac{5}{4a_{I+1}}-
\frac{4(a_I^2+3a_Ia_{I+1}+a_{I+1}^2)}{(a_I+a_{I+1})^3}\,,
\\ & &\int_0^\infty d\xi 
\left(H'-\frac{H}{\xi}\right)^2=
\frac{1}{4\alpha}\,, \\
& &\int_0^\infty d\xi
\frac{1}{\xi^2}K_1^2H^2=
\frac{\alpha^2(56a_1^4+132a_1^3\alpha+111a_1^2\alpha^2+39a_1\alpha^3+5\alpha^4)}%
{4a_1(a_1+\alpha)^3(2a_1+\alpha)^3}\,,
\\ & &\int_0^\infty d\xi
\xi^2\left(\frac{H^2}{\xi^2}-1\right)^2=
\frac{635}{864\alpha^3}\,.
\end{eqnarray}
Therefore, the energy expressed by the variational parameters becomes
\begin{eqnarray}
& &E_N=\frac{4\pi v}{g}\left[\frac{41}{64}\sum_{I=1}^{N+1}a_I+
\frac{f^2}{v^2}\sum_{I=1}^N\left\{\frac{5}{4a_I}+\frac{5}{4a_{I+1}}-
\frac{4(a_I^2+3a_Ia_{I+1}+a_{I+1}^2)}{(a_I+a_{I+1})^3}\right\}\right.\nonumber \\
& &\qquad\quad\quad+\left.\frac{1}{8\alpha}+
\frac{\alpha^2(56a_1^4+132a_1^3\alpha+111a_1^2\alpha^2+39a_1\alpha^3+5\alpha^4)}%
{4a_1(a_1+\alpha)^3(2a_1+\alpha)^3}+
\frac{635\lambda}{3456g^2\alpha^3}\right]\,.
\label{ven}
\end{eqnarray}

\begin{figure}[ht]
\centering
\includegraphics
{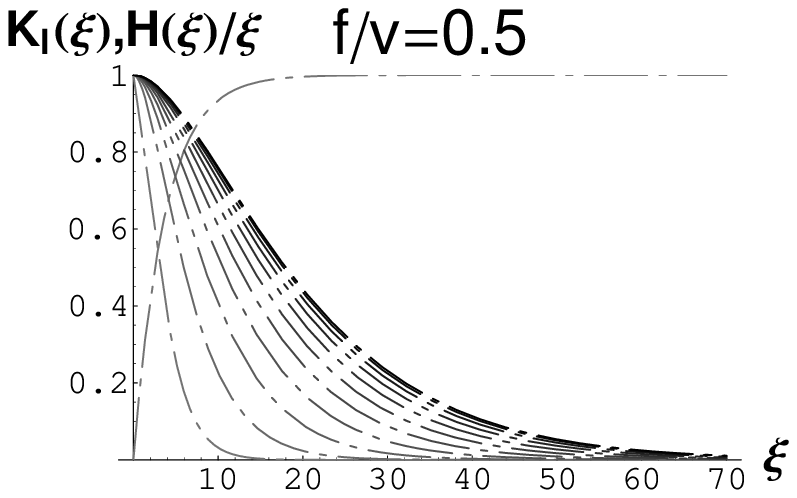}
\includegraphics
{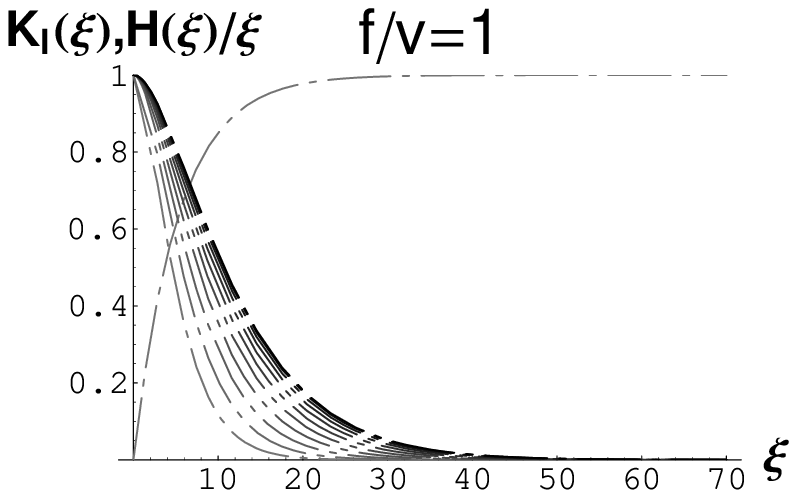}\\
\vspace{1cm}
\includegraphics
{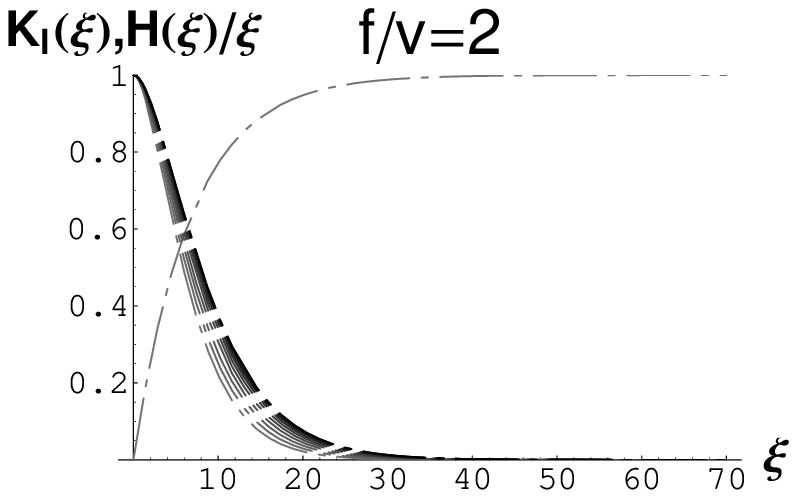}
\caption{$K_I$ and $H/\xi$ are shown for $N=10$, $\lambda=0$.
In each graph, $K_I<K_{I+1}$ $(1\le I\le N=10)$ at any $\xi$.
The three graphs correspond to the cases of $f/v=0.5$,
$f/v=1$, and $f/v=2$, respectively.
}
\label{fig2}
\end{figure}

We evaluate the minimum value of this energy by numerical calculation with
\textit{Mathematica} \cite{wolfram}.
Thus, we get the approximate solution of $K_I(\xi)$ and $H(\xi)/\xi$,
and the case of $N=10$ and $\lambda=0$ is shown in Fig.~\ref{fig2}.

The region of non-vanishing $K_I$ can be interpreted as
the region where the $I$-th massive vector bosons ($a=1,2$) condensate.
For larger values of $f/v$, the ranges of finite $K_I$ become narrower
and degenerate, while the distance where $H/\xi\sim 1$ becomes larger.

We obtain the energy of the monopole in the limiting case $\lambda/g^2=0$
for the cases where $N=0, 1, 5, \mathrm{and~} 10$ and $f/v=0.5, 1, \mathrm{and~}2$.
\begin{eqnarray}
E_0&=&\frac{4\pi v}{g}\times 1.05\cdots\,,\quad (\mbox{BPS monopole})\\
E_1&=&\frac{4\pi v}{g}\times 1.41\cdots\,, \quad (f/v=0.5)\\ E_1&=&\frac{4\pi
v}{g}\times 1.47\cdots\,, \quad (f/v=1)\\ E_1&=&\frac{4\pi v}{g}\times
1.48\cdots\,, \quad (f/v=2)\\ E_5&=&\frac{4\pi v}{g}\times 2.07\cdots\,,  \quad
(f/v=0.5)\\ E_5&=&\frac{4\pi v}{g}\times 2.37\cdots\,,  \quad (f/v=1)\\
E_5&=&\frac{4\pi v}{g}\times 2.51\cdots\,,  \quad (f/v=2)\\
E_{10}&=&\frac{4\pi v}{g}\times 2.47\cdots\,, \quad (f/v=0.5)\\
E_{10}&=&\frac{4\pi v}{g}\times 2.99\cdots\,, \quad (f/v=1)\\
E_{10}&=&\frac{4\pi v}{g}\times 3.32\cdots\,, \quad (f/v=2)
\end{eqnarray}
From these results, we roughly estimate that the energy of the monopole
($\lambda=0$) is
\begin{equation}
E_N\sim\frac{4\pi v}{g}\sqrt{N+1}\quad (\lambda=0)\,,
\end{equation}
since the difference that appears due to different $f/v$ is smaller than that due
to different $N$.
We find that our approximate values of the static energies for
$\lambda=0$ are well fitted to $E_N\approx\frac{4\pi v}{g}\times 1.94\times W(0.62
N+0.96)$, where $W(x)$ is the Lambert $W$-function, which is slightly smaller than
$\frac{4\pi v}{g}\sqrt{1+N}$ for large
$N$.
This is in contrast to the rather large dependence of the
profiles of solutions for $K_I$ and $H/\xi$ on $f/v$ (Fig.~\ref{fig2}).

On the other hand, we know the exact value of the energy of the BPS
limit \cite{BPS1,BPS2} for the `t Hooft--Polyakov monopole, which corresponds
to $E_0$ for $\lambda=0$, as
\begin{equation}
E(\lambda=0)=\frac{4\pi v}{g}\,.
\end{equation}
Comparing these values, we find that the energy of the BPS monopole in the
multi-vector boson theory is obtained by replacing $g\rightarrow g/\sqrt{N+1}$
in that of the usual BPS monopole.

Note that we only show the case of $\lambda\rightarrow0$.
However, we confirmed that the energy of the monopole changes at most factor two
for a finite value of $\lambda/g^2$ in general. 

\section{Magnetic charge of the monopole}
\label{sec6}

In this section, we specify the magnetic charge of the monopole in the
multi-vector boson theory obtained in the previous section.
First of all, we should discuss the definition of electric charge. 
As in Section \ref{sec3}, if we choose $\phi^3=v$, the massless gauge field
satisfies
\begin{equation}
A_{1\mu}^3=A_{1\mu}^3=\cdots=A_{N+1,\mu}^3\equiv\frac{1}{\sqrt{N+1}}A_\mu^3\,.
\end{equation}
The normalization factor is determined by the canonical form of the Lagrangian
density of this zero-mode field.
Therefore, if the charged matter field is virtually coupled only to
$A_{1\mu}$, similar to that in the triplet Higgs field, the electric charge of the
matter field $e$ becomes
\begin{equation}
e=\frac{g}{\sqrt{N+1}}\,,
\end{equation}
and the field strength satisfies
$G_{1\mu\nu}^3=G_{1\mu\nu}^3=\cdots=G_{N+1,\mu\nu}^3\equiv
\frac{1}{\sqrt{N+1}}G_{\mu\nu}^3$.

Now, we consider the magnetic field far from the monopole.
The projection of the vacuum expectation values of the field strength
\cite{GO,Rubakov,Weinberg} is
\begin{equation}
\lim_{r\rightarrow\infty}F_{ij}=\lim_{r\rightarrow\infty}
\hat{\phi}^aG_{ij}^a=\lim_{r\rightarrow\infty}
\frac{1}{\sqrt{N+1}}\sum_{I=1}^{N+1}\hat{\phi}^aG_{Iij}^a
=\frac{\sqrt{N+1}}{g}
\left(-\varepsilon_{aij}\frac{x^a}{r^3}\right)\,,
\end{equation}
where $\hat{\phi}^a=\phi^a/v$. Then, the magnetic field $B^i$ is asymptotically
\begin{equation}
B^i=-\frac{\sqrt{N+1}}{g}\frac{x^i}{r^3}=-\frac{x^i}{er^3}\,.
\end{equation}
Comparing this magnetic field representation with the magnetic field created by a
point magnetic charge $g_m$ 
\begin{equation}
B^i=\frac{g_m}{4\pi}\frac{x^i}{r^3}\,,
\end{equation}
the magnetic charge $g_m$ of our monopole is
\begin{equation}
g_m=-\frac{4\pi}{g}\sqrt{N+1}=-\frac{4\pi}{e}\,.
\end{equation}
This relation is the same as that for the `t Hooft--Polyakov monopole.

The static energy of the monopole in the multi-vector boson theory that was
described in the previous section can be rewritten as
\begin{equation}
E_N\sim\frac{4\pi v}{g}\sqrt{N+1}=\frac{4\pi v}{e}\,,
\end{equation}
which is the same as the mass of the `t Hooft--Polyakov monopole (or, the case
of $N=0$).

\section{Summary and discussion}
\label{sd}

In this paper, we studied the static, spherically symmetric monopole
solutions in the multi-vector boson theory with $N+1$ sets of vector
bosons with the gauge coupling $g$. The theory includes two mass scales $f$ and
$v$. 
We found that $3N+2$ massive vector bosons and a single massless vector boson (of
the electromagnetic field) appear according to the theory described in
Sec.~\ref{sec4}.  We used a simple variational method to obtain approximate
solutions in Sec.~\ref{sec5}. 
The solution of $K_I$ shows that the regions of existence of massive vector fields
have a multi-layer structure, where massive bosons `stratify'.
Although the profile of condensation of the massive degrees of
freedom is sensitive with respect to both $N$ and $f/v$,
the mass of the monopole is approximately $E_N\sim\frac{4\pi
v}{g}\sqrt{N+1}=\frac{4\pi v}{e}$, where $e$ is the electric charge defined in the
theory. It is necessary to conduct a more accurate investigation for
obtaining the precise dependence of mass of the monopole on $N$.

The model used in this study is the simplest one; therefore, we would like to
investigate more general models, which have different coupling constants for
different gauge fields or have complicated mass matrices as in the clockwork theory
\cite{CI,KaRa,GM1,FPRT,KeRi,HTT}.

Another possible connection to a phenomenological model can be considered in a
model with symmetry breakdown by a Higgs doublet, as in the standard model.
In 1997, Cho and Maison \cite{CM} found an electroweak monopole solution in the
standard model. The Cho--Maison monopole and its generalization have been studied 
further \cite{KYC,CKY}, and an experimental search for them is going on
\cite{EMY,MoEDAL}. We wish to investigate the multi-vector boson theory with a
doublet Higgs field and compare the properties of its monopoles with those of the
Cho--Maison monopoles.

We also wish to study a scenario in which the monopoles in the multi-vector boson
theory represent the dark matter in the universe. 
Since the present model of multi-vector boson theory has two symmetry breaking
scales
$f$ and $v$, and there can be various mass spectra of massive vector bosons as
seen in Sec.~\ref{sec4}, we need to perform a detailed study on the process of
symmetry breaking and (time-dependent) monopole production. 

\acknowledgments
We thank
Hideto Manjo for useful comments on numerical estimations.

\section*{Conflict of Interests}
The authors declare that there is no conflict of interests regarding the
publication of this paper.

\bibliographystyle{apsrev4-1}

\end{document}